\documentclass[12pt,preprint]{aastex}
\usepackage{lineno}

\newcommand{\be}{\begin{equation}}
\newcommand{\ee}{\end{equation}}
\def\lta{\,\raise 0.3 ex\hbox{$ < $}\kern -0.75 em
 \lower 0.7 ex\hbox{$\sim$}\,}
\def\gta{\,\raise 0.3 ex\hbox{$ > $}\kern -0.75 em
 \lower 0.7 ex\hbox{$\sim$}\,} 
\newcommand{\texp}{t_{\rm exp}}

\newcommand{\trunk}{r_{\scriptstyle X}} 
 
\newcommand{\heat}{{\cal H}} 
\newcommand{\conduct}{K_{\rm c}} 
\newcommand{\mplan}{M_{\rm plan}} 
\newcommand{\rplan}{R_{\rm plan}} 
\newcommand{\crlum}{L_{\rm p}} 
\newcommand{\ndot}{{\dot {\cal N}}} 
\newcommand{\sigmabar}{\langle\sigma\rangle} 
\newcommand{\radioactive}{P_{\scriptscriptstyle SLR}} 
\newcommand{\rxcon}{\omega_{\rm x}} 

\begin{document} 

\title{Radioactive Planet Formation}  

\author{Fred C. Adams$^{1,2}$}

\affil{$^1$Physics Department, University of Michigan, Ann Arbor, MI 48109} 
\affil{$^2$Astronomy Department, University of Michigan, Ann Arbor, MI 48109}

\begin{abstract}

Young stellar objects are observed to have large X-ray fluxes and are
thought to produce commensurate luminosities in energetic particles
(cosmic rays).  This particle radiation, in turn, can synthesize
short-lived radioactive nuclei through spallation.  With a focus on
$^{26}$Al, this paper estimates the expected abundances of radioactive
nulcei produced by spallation during the epoch of planet formation.
In this model, cosmic rays are accelerated near the inner truncation
radii of circumstellar disks, $\trunk\approx0.1$ AU, where intense
magnetic activity takes place.  For planets forming in this region,
radioactive abundances can be enhanced over the values inferred for
the early solar system (from meteoritic measurements) by factors
of $\sim10-20$. These short-lived radioactive nuclei influence the
process of planet formation and the properties of planets in several
ways. The minimum size required for planetesimals to become
fully molten decreases with increasing levels of radioactive
enrichment, and such melting leads to loss of volatile components
including water. Planets produced with an enhanced radioactive
inventory have significant internal luminosity which can be comparable
to that provided by the host star; this additional heating affects
both atmospheric mass loss and chemical composition.  Finally, the
habitable zone of red dwarf stars is coincident with the magnetic
reconnection region, so that planets forming at those locations will
experience maximum exposure to particle radiation, and subsequent
depletion of volatiles.

\end{abstract} 

$\,\,\,$ {\it UAT Concepts:} Exoplanet formation (492); 
Exoplanet astronomy (486) 

\section{Introduction} 
\label{sec:intro} 

The collection of observed planets orbiting other stars shows enormous
diversity, which poses a fundamental problem for understanding the
process of planet formation. One component of the explanation is that
planets are likely to form with a variety of background conditions,
which are provided by the circumstellar disks that give rise to these
companions. This paper explores one aspect of the problem by
considering the range of possible abundances of short-lived
radionuclides (SLRs) that are present during the formation and early
evolution of planetary bodies.  More specifically, we consider the
possible range of nuclear enrichment resulting from spallation due to
cosmic rays produced by the central stars, with a focus on the isotope
$^{26}$Al.  Young stellar objects --- both protostars and
pre-main-sequence stars --- are observed to have substantial X-ray
luminosities \citep{feigelmont}, which is thought to coincide with
comparable emission levels of energetic particles (cosmic rays). In
the immediate vincinity of the star, this intense particle radiation
can drive nuclear reactions through spallation and produce substantial
quantities of radioactive nuclei.\footnote{For completeness we note
  that `spallation' refers to nuclei being broken apart, whereas the
  production of $^{26}$Al does not always break apart the target
  nuclei.  Nonetheless, this paper follows standard usage in the
  astronomical literature where the term spallation refers to all
  nuclear reactions driven by particle radiation such as cosmic rays.}

In this context, the short-lived radioactive nuclei of interest have
lifetimes of order $\sim1$ Myr and can be produced with sufficient
adundance to influence planetary properties. For example,  
radioisotopes of potential interest include $^{10}$Be, $^{26}$Al,
$^{36}$Cl, $^{41}$Ca, $^{53}$Mn, and others, which are thought to have
been present in unexpectedly large numbers in the early formative
stages of our own solar system (\citealt{lee1977};
cf. \citealt{jura2013}).

Radioactive nuclei, if present in sufficient abundances, can provide
an important energy source during the planet formation process (see
\citealt{urey1955} to \citealt{reiter2020}).  Sufficiently large
planetesimals will melt due to their internal energy supply
\citep{schramm1971,hevey2006}, which leads to the differentiation of
the rocky bodies \citep{wasserburg,moskogaidos}.  After their
formation is complete, the planets will have an additional luminosity
source due to internal radioactivity. This additional heating can
remove volatile components from both the raw materials and the planets
themselves \citep{grimm1993,ikoma2018}. More specifically, radioactive
heating acts to dehydrate planetesimals \citep{lichtenberg} and thus
affects the water content of forming planets. Radioactive nuclei also
provide a significant source of ionization (especially in the outer 
disk; \citealt{cleeves2013}), which in turn affects both the chemical
composition of the disk and possiblity of driving disk accretion
through the magneto-rotational-instability.

A great deal of previous work has focused on the radioactive
enrichment of our solar system, where meteoritic evidence indicates
that the early solar nebula was enriched in SLRs relative to
abundances expected in the interstellar medium
\citep{lee1977,wasser1985}. Such considerations place interesting
constraints on the birth environment of the Sun \citep{adams2010}.
One of the challenges facing such theoretical explanations is that we
observe enhancements in several nuclear species, including $^{10}$Be,
$^{26}$Al, $^{41}$Ca, $^{60}$Fe, $^{53}$Mn, $^{107}$Pd, and $^{129}$I,
and the relative abundances of the various nuclei must be accounted
for (compare \citealt{shu1997,lee1998,gounelle2001,goswami2001,leya2003,
duprat2007,liu2009,desch2010,gounelle2013,sossi2017,jacquet2019}).  In 
the present application, however, we are primarily concerned with the
nuclear species that provide the most significant source(s) of
ionization and heating. The expected time scales for both planet
formation and disk evolution fall in the range 1 -- 10 Myr, so that
nuclei with half-lives longer than $\sim10$ Myr are of less interest. 
Since $^{60}$Fe must be produced via stellar nucleosynthesis, it is
not considered here. The abundances of other nuclear species are
relatively low, with the mass fraction of $^{26}$Al larger than
$^{36}$Cl by a factor of $\sim4$ and larger than $^{53}$Mn by a factor
of $\sim10$, so that the isotope $^{26}$Al is likely to provide the
largest impact. This treatment thus focuses on the production of
$^{26}$Al (but other isotopes that can be synthesized through
spallation can be scaled from these results).

In this paper, we present a scenario for the production of cosmic
radiation in the inner regions of disks associated with young stellar
objects (Section \ref{sec:produce}). Through spallation reactions,
this particle radiation can enrich the rocky material that eventually
forms planets, where the enrichment levels are estimated in Section
\ref{sec:exposure}. For planets forming in disk regions near
$r\sim0.1$ AU, we find that the abundances of SLRs can be significantly
enhanced relative to those inferred for our solar system (which are
already thought to be relatively large). This additional source of
heating and ionization can affect young and forming planets, as
briefly outlined in Section \ref{sec:effects}. The paper concludes in
Section \ref{sec:conclude} with a summary of results and a discussion
of their implications.

\section{Production of Particle Radiation} 
\label{sec:produce} 

In the astrophysical scenario considered here, the production of
particle radiation (cosmic rays) takes place near the magnetic
truncation radius of the inner disk. In this region, the magnetic
fields experience reconnection events which act to accelerate
particles to high energies, thus creating a local flux of cosmic rays.
This section outlines the basic properties for this source of particle
radiation (for further detail, see
\citealt{shu1996,shu1997,lee1998,shu2001,gounelle2001}; 
\citealt{leya2003,duprat2007}; and references therein). 

Young stellar objects, both protostars and T Tauri stars, are observed
to have strong magnetic fields with typical surface values of order
$B_\ast\approx1-2$ kilogauss \citep{jkrull}. Near the star, these
fields have enough pressure to control the inward accretion flow from
the disk. As a result, the disk is truncated at the radius where the
field pressure balances the ram pressure due to accretion. The
resulting truncation radius $\trunk$ can be written in the form 
\be
\trunk = \rxcon \left( {B_\ast^4 R_\ast^{12} \over 
G M_\ast {\dot M}^2} \right)^{1/7} \,,
\label{trunk} 
\ee
where $B_\ast$ is the field strength on the surface of the star,
$R_\ast$ is the stellar radius, $M_\ast$ is the stellar mass, and
${\dot M}$ is the mass accretion rate (from the disk onto the star, 
at the truncation point). The general form of this expression follows
from dimensional analysis, whereas the dimensionless parameter
$\rxcon$ varies with the details of the model
\citep{ghoshlamb,blandford,najita}, but is expeced to be of order
unity. We are also implicitly assuming that the stellar magnetic field
has a substantial dipole component. 

\begin{figure}
\includegraphics[scale=0.70]{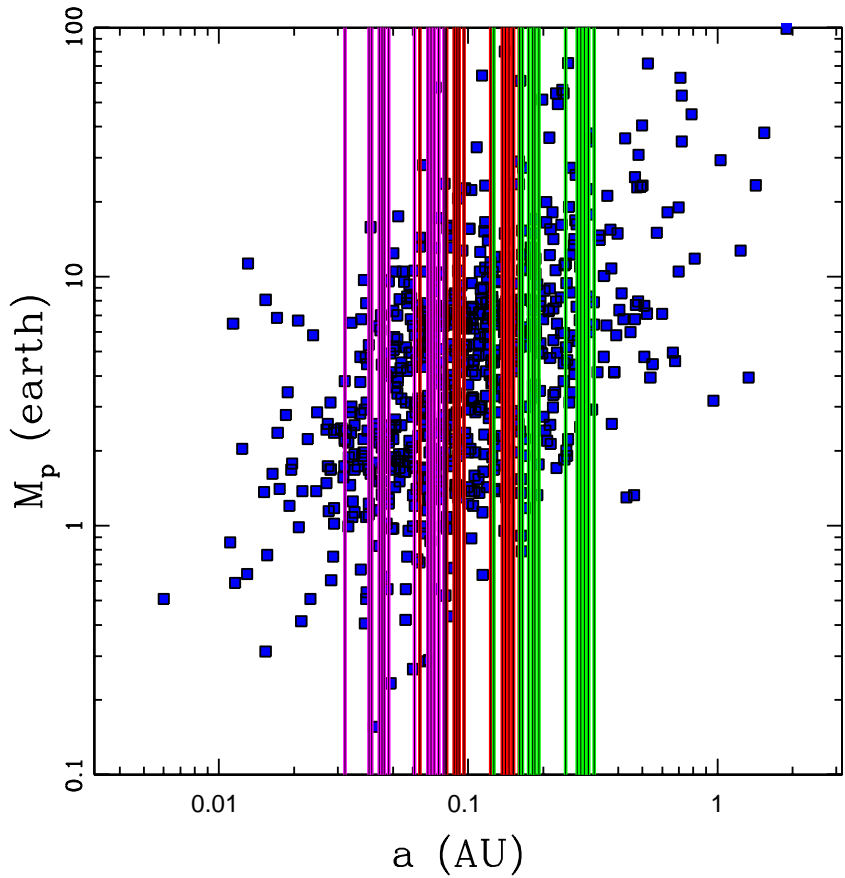}
\vskip-1.50truein
\caption{Estimated locations of magnetic truncation radius in observed
T Tauri star/disk systems and the exoplanet population. The background 
points depict observed exoplanets in the current data base for systems
containing 3 or more planets. The red vertical lines show the
estimated location of the magnetic truncation radius for a sample of
15 T Tauri stars with observed magnetic field properties. Additional 
lines show magnetic truncation radii that are a factor of two larger 
(green) and smaller (magenta) than the fiducial values. } 
\label{fig:rmag} 
\end{figure}

In this scenario, mass accretes through the circumstellar disk and
flows inward to the truncation radius given by equation (\ref{trunk}),
where the magnetic pressure becomes larger than the ram pressure of
the flow.  Gas parcels can accrete along the field lines and
eventually reach the stellar surface, whereas rocky bodies of
sufficient size are left behind in the disk region surrounding the
radius $\trunk$. The magnetic coupling between the star and disk acts
to enforce a time-averaged rotation rate for the star that matches the
Keplerian rotation rate at $\trunk$. Both the accretion flow and the
magnetic field lines are subject to fluctuations. As a result, the
magnetic field in the truncation region is expected to wind up, and
then reconnect, and thereby produce stressed fields and sporadic
energy release, which in turn leads to both X-ray emission and
particle acceleration. The net result of this star/disk coupling is
the production of a particle luminosity $\crlum$ that is a substantial
fraction of the photon luminosity $L_\ast$ \citep{lee1998}. Moreover,
$\crlum$ is is expected to be comparable to (but somewhat less than)
the X-ray luminosity $L_X$ \citep{padovani2016}, which has been
observed for a collection of young stellar objects
\citep{feigelson,preibisch}. As a result, we expect the cosmic ray
luminosity, the X-ray luminosity, and the stellar luminosity to obey
the ordering relation 
\be 
\crlum \sim 10^{-1} L_X \sim 10^{-4} L_\ast \,.
\label{luminosity} 
\ee
The stellar luminosity $L_\ast\sim1L_\odot$ for typical pre-main
sequence stars (but is somewhat larger for protostellar objects).
Note that lower mass stars have smaller main-sequence luminosities,
but evolve more slowly via PMS contraction, and these effects tend 
to cancel out. Finally, we note that the particle luminosity 
$\crlum$ includes protons, alpha particles, and $^3$He nuclei. 

In this model, the cosmic ray luminosity is generated in the disk
region near the truncation radius $\trunk$. Here we assume that the
annulus defined by $\trunk/2\lta{r}\lta\trunk$ provides the bulk of
the cosmic ray flux (see \citealt{najita,shu1997}). Furthermore,
the mass accretion rate and other parameters appearing in equation
(\ref{trunk}) vary with time, so that the outer boundary (at $\trunk$)
moves inward and outward, perhaps varying by a factor of 2. This
annular region is coincident with that of X-ray emission, where
observations suggest that the plasma in the immediate vincinity of the
reconnection events has a column density of order $\sim10^{-4}$ g
cm$^{-2}$ \citep{shu1997,lee1998}. This low column density implies 
that the cosmic rays flux produced alongside the X-rays will not 
be highly attenuated before interacting with rocky material in the 
region. 

Within the reconnection annulus, cosmic rays are tied to the local
magnetic fields, which are twisted and chaotic. Particle propagation
is controlled by two important length scales. The first is the scale
height of the inner disk, $H\sim\trunk/20$, which determines the
typical length scale of the tangled magnetic field. The second is the
magnetic gyroradius $r_g=\gamma mcv/qB$, where $r_g\ll{H}$ for the particle
energies and field strengths of interest.  The following picture
emerges: The cosmic rays are free to move along the field lines, which
are highly convoluted with step length $\sim H$. The particles are
closely tied to the field lines and thus experience little
perpendicular propagation. In order to escape the reconnection region,
the cosmic rays must travel a net distance of order $\trunk$. Since
the particles execute a random walk with step length $H$, the total
distance $d$ traveled to escape is given by $d\sim\trunk^2/H$
$\sim20\trunk$. As the cosmic rays travel through such a long
distance, and cross the midplane of the disk multiple times, a large
fraction of the particles will interact with rocky material located
within the annulus.

The cosmic ray luminosity $\crlum$ corresponds to all particles with
energy above some minimum value, $E\ge E_0$, where $E_0\sim10$ MeV.  
The cosmic rays with energy above this benchmark value are expected 
to have a power-law spectrum, which can be written in the form
\be
{df\over dE} = {p-1 \over E_0} 
\left({E \over E_0}\right)^{-p} \,,
\label{crspectrum} 
\ee
where the index $p$ depends on the source
\citep{hollebeke,reames1997,reames1999,mewaldt2012,reames2013}.
For cosmic radiation emitted by the Sun through the action of gradual
flaring activity, the power-law index $p\approx2.7$. The Sun also
accelerates cosmic rays through the action of impulsive flares, which
result in a steeper index $p\approx3.5$. Galactic cosmic rays have a
similar power-law index at high energies. Given this range of
possibilities, for the sake of definiteness we adopt $p=3$ for the
estimates of this paper. Note the the distribution of equation
(\ref{crspectrum}) is normalized such that $\int_{E_0}^\infty
(df/dE)dE=1$.

The cosmic ray luminosity $\crlum$ is defined in terms of the energy
emitted in particles. In order to determine the reaction rates for
spallation, we also need to determine the rate $\ndot$ of particle
emission. With the cosmic ray spectrum (\ref{crspectrum}) specified,
we can define the luminosity density $L_E$ in cosmic rays and the
corresponding quantity $\ndot_E$ for particle number according to 
\be
L_E = \crlum {p-2\over E_0} \left({E \over E_0}\right)^{1-p} 
\qquad {\rm and} \qquad \ndot_E = {\crlum (p-2) \over E_0^2} 
\left({E \over E_0}\right)^{-p} \,. 
\ee
These distributions are normalized such that 
\be
\int_{E_0}^\infty L_E \, dE = \crlum \,, \qquad 
\int_{E_0}^\infty \ndot_E \, dE = \ndot  \,,
\qquad {\rm and} \qquad 
\ndot = {p-2\over p-1} {\crlum \over E_0} \,. 
\ee
We can also specify the particle luminosity and particle emission rate
for cosmic rays with energy above a threshold value, $E>E_{th}>E_0$, 
\be
L(E>E_{th}) = \left( {E_0 \over E_{th}}\right)^{p-2} \crlum 
\qquad {\rm and} \qquad 
\ndot(E>E_{th}) = \left( {E_0 \over E_{th}}\right)^{p-1} \ndot \,. 
\label{threshlum} 
\ee

In addition to protons, the locally produced cosmic ray luminosity
contains alpha particles ($^4$He nuclei), helions ($^3$He nuclei), and
a small fraction of larger nuclei. For the sake of definiteness, we
assume here that the values for $\ndot$ given above correspond to the
particle output in protons and alpha particles, which make up the
majority of particles accelerated from the reconnection region. We
further assume that the emission rate for alpha particles is smaller
by an order of magnitude, so that $\ndot_\alpha=0.1\ndot$ and
$\ndot_p=0.90\ndot$. This assumption is consistent with both the
cosmic abundance of helium and that content of galactic cosmic rays.
We then consider an additional admixture of helions $h$ ($^3$He),
where the relative number of particles is more uncertain. For galactic
cosmic rays, the ratio $[h/\alpha]$ is typically $\sim1/10$
\citep{formato}.  On the other hand, the ratio $[h/\alpha]$ can
approach and even exceed unity in the most energetic solar flares
\citep{garrard1973,fisk1978}. Here we adopt an intermediate value by
considering the additional luminosity of helions to be given by
$\ndot_h=0.03\,\ndot$.

The magnetic field strengths and other parameters appearing in
equation (\ref{trunk}) have been measured for a collection of T Tauri
stars \citep{jkrull}, allowing us to estimate the expected values of
the truncation radius $\trunk$. The resulting values are shown in
Figure \ref{fig:rmag}, where they are plotted alongside the observed
exoplanets found in multi-planet systems (see \citealt{batalha} and
subsequent work), where estimates of the planet masses are plotted
as a function of semimajor axis. The vertical red lines depict the
locations of the radii $\trunk$ for the 15 T Tauri stars in the
observed sample and for dimensionless parameter $\rxcon=1$. Since the
mass accretion is episodic, and its rate ${\dot M}$ can vary by an
order of magnitude, the truncation radius $\trunk$ can vary by (at
least) a factor of two.  Other parameters (including $\rxcon$) can
also vary. To account for this diversity, Figure \ref{fig:rmag} also
shows the locations of the truncation radii that are a factor of two
larger (green lines) and a factor of two smaller (magenta lines) than
their fiducial values. The locations of the magnetic truncation radii
thus coincide with the radial locations of observed exoplanets. Since
the particle radiation (outlined above) originates within this same
region, this class of exoplanets will be exposed to intense cosmic ray
radiation during their early evolutionary stages.

The young stars used to determine the magnetic truncation radii in
Figure \ref{fig:rmag} and the (older) host stars of the planetary
systems have comparable properties. Specifically, the sample of T
Tauri stars \citep{jkrull} has a distribution of masses that can be
characterized by $M_\ast=0.72\pm0.45M_\odot$ (in the form of mean
$\pm$ standard deviation).  The host stars for the planetary sample
shown in the figure have comparable masses, which fall in the range
$M_\ast=0.10-1.35M_\odot$.  The resulting values for the magnetic
truncation radii are $\trunk=0.11\pm0.034$ AU. Finally, the
corresponding X-ray luminosities are also measured for the T Tauri
sample, where $L_X = 1.53\pm1.32 \times 10^{-3} L_\odot$. These values
define a characteristic particle luminosity $\crlum$ = 0.1
$L_X\approx1.5\times10^{-4}L_\odot$, which corresponds to
$\ndot=3.75\times10^{34}$ particle/sec.

\section{Irradiation of Rocky Targets by Local Cosmic Rays} 
\label{sec:exposure} 

The previous section outlined the model for the generation of cosmic
rays in the annular region near the truncation radius $\trunk$.  For
planets forming in this region, this section estimates how spallation 
reactions can enrich the raw materials with radioactive nuclei. The 
spallation reactions act on target nuclei and generate radioactive 
isotopes. The reactions of interest have the general form 
\be
\chi + T \longrightarrow \radioactive + \xi_j
\ee
where $T$ is the target nucleus, $\radioactive$ is the radioactive
product, and $\chi$ is the projectile (a proton $p$, alpha particle
$\alpha$, or helion $h$). The $\xi_j$ represent the additional
particles that are produced, generally protons $p$, neutrons $n$, and
alphas $\alpha$ with varying numbers per reaction. Here we focus on
the radioactive products $\radioactive\sim$ $^{26}$Al.

The target nuclei of interest generally reside in rocky bodies of
various sizes.  In the current paradigm of planet formation, dust
grains (with characteristic size $R\sim0.1-1\mu$m) grow into larger
rocky bodies known as pebbles (with characteristic size $R\sim0.1-1$
cm). After the concentration of pebbles becomes large enough, a
streaming instability \citep{youdin2005,drazkowska} or some other
process consolidates the pebbles into much larger bodies called
planetesimals with characteristic size $R\sim10-100$ km \citep{simon}.
These planetesimals subsequently accummulate into full-sized
planets. This paper focuses on the formation of rocky planets with
masses in the range $\mplan$ = $1-10M_\oplus$. Larger planets can
sometimes form, where rocky cores of mass $M_c\sim10M_\oplus$ accrete
gaseous envelopes and reach Jovian masses. However, such hot Jupiters
are relatively rare in the reconnection region, and are not addressed
in this work.

\subsection{Cosmic Ray Exposure in the Optically Thin Regime} 
\label{sec:opthin} 

In the scenario under consideration (Section \ref{sec:produce}),
magnetic fields from the star truncate the circumstellar disk at
radius $\trunk$ and cosmic rays are generated in the surrounding
region. The resulting number flux $\Phi_{CR}$ in cosmic rays in 
the reconnection annulus is thus given by  
\be
\Phi_{CR} \approx {\ndot \over \pi \trunk^2} \, ,
\ee
where $\ndot$ is the total number of cosmic rays emitted per unit 
time. 

We first consider a particular spallation reaction operating in the
optically thin regime. The rate $\Gamma$ at which a target nucleus
absorbs cosmic rays, and thereby produces a radioactive isotope, can
be written in the form
\be
\Gamma_k = {1 \over \pi \trunk^2} 
\int_{E_0}^\infty \ndot_E \sigma_k(E) dE
\equiv {\ndot_k \sigmabar_k \over \pi \trunk^2} \,,
\ee
where $\sigma_k(E)$ is the cross section for the reaction of interest
and $\sigmabar_k$ is the cross section averaged over the distribution
of cosmic ray energies. The index $k$ labels the reaction of interest.  
Note that the total cosmic ray luminosity $\ndot$ (in particle number)
includes projectiles of all types, whereas the only a fraction
$\ndot_k$ participate in the reaction.

In this application, we are interested in the amount of $^{26}$Al
produced, which can be synthesized through a number of spallation
reactions. Specifically, reactions involving targets of $^{27}$Al,
$^{26}$Mg, and $^{28}$Si are important for proton spallation, and
targets of $^{26}$Mg and $^{28}$Si are relevant for alpha particle
radiation. Targets of $^{24}$Mg, $^{25}$Mg, $^{27}$Al, and $^{28}$Si
are relevant for spallation with helion ($^3$He) particles. These
reactions are summarized in Table \ref{table:react}, along with 
defining parameters of interest (defined below). 

The abundance of $^{26}$Al is measured relative to a reference
isotope, taken here to be $^{27}$Al. Over a given exposure time
$\texp$, the abundance ratio is given by the sum 
\be
{\cal R}_{26} = {X_{26} \over X_{27}} \approx 
{\ndot \texp \over \pi \trunk^2} \sum_k \sigmabar_k 
{X_k \over X_{27}} {\ndot_k \over \ndot} \,. 
\label{ratone} 
\ee
The first factor in the sum accounts for the cross section of the
interaction, the second factor is the ratio of abundances of the
target nuclei, and the third factor is the ratio of abundances of the
projectiles. 

Note that the above treatment represents an upper limit on the
expected abundance ratio. The estimate does not include the back
reaction on the SLR production rate due to the loss of target nuclei
or the decay of the radioactive products.  If the value of 
${\cal R}_{26}$ approaches unity, then the ratio will be smaller than
assumed here due to the decline in the supply of targets, although
this correction is small. In additon, however, if the exposure time
$\texp$ become comparable to the decay time of the SLR, then the ratio
of SLRs will be smaller due to radioactive decay (which occurs at the
rate $\lambda=\ln2/t_{1/2}$). Taking into account both of these
complications, we can generalize the result (\ref{ratone}) to obtain 
\be
{\cal R}_{26} = 
{\ndot \sigmabar_T \over \pi \trunk^2} {1 \over \lambda - \Gamma} 
\left[ 1 - {\rm e}^{-(\lambda-\Gamma)\texp} \right] 
\rightsquigarrow
{\ndot \sigmabar_T \over \pi \trunk^2 \lambda} \,, 
\label{rattwo} 
\ee
where we have defined a total effective cross section 
\be
\sigmabar_T \equiv \sum_k \sigmabar_k 
{X_k \over X_{27}} {\ndot_k \over \ndot} \,. 
\label{sigtotal} 
\ee
Note that we expect $\lambda\gg\Gamma$. In the limit of short exposure
times, specifically $\texp\ll t_{1/2}$, we recover equation
(\ref{ratone}).  In the opposite limit where $\texp\gg t_{1/2}$, we
find that ${\cal R}_{26}\to\Gamma/(\lambda-\Gamma)\approx\Gamma/\lambda$.  
In other words, the effective exposure time becomes
$\texp\sim\lambda^{-1}=t_{1/2}/\ln(2)$.

\bigskip 

\begin{table}[h] 
\centerline{\bf Nuclear Reactions for Spallation}
\vskip8pt
\begin{tabular}{lccc} 
\hline
\hline
Reaction & Cross Section & Target Abundance & Projectile Fraction \\ 
$\,$ & $\langle\sigma\rangle$ (mb) & $X_T/X_H$ & $\ndot_\chi/\ndot$ \\ 
\hline \hline 
$^{26}$Mg$(p,n)^{26}$Al & 417 & $4.6\times10^{-6}$ & 0.9 \\ 
$^{27}$Al$(p,pn)^{26}$Al & 44 & $3.5\times10^{-6}$ & 0.9 \\ 
$^{28}$Si$(p,2pn)^{26}$Al & 7.9 & $3.8\times10^{-5}$ & 0.9 \\
\hline 
$^{24}$Mg$(\alpha,pn)^{26}$Al & 30 & $3.3\times10^{-5}$  & 0.1 \\ 
$^{28}$Si$(\alpha,\alpha pn)^{26}$Al & 4.4 & $3.8\times10^{-5}$ & 0.1 \\ 
\hline
$^{24}$Mg($^{3}$He,$p)^{26}$Al & 216 & $3.3\times10^{-5}$ & 0.03 \\ 
$^{25}$Mg($^{3}$He,$pn)^{26}$Al & 316 & $4.2\times10^{-6}$ & 0.03 \\ 
$^{27}$Al($^{3}$He,$\alpha)^{26}$Al & 33 & $3.5\times10^{-6}$ & 0.03 \\ 
$^{28}$Si($^{3}$He,$p\alpha)^{26}$Al & 60 & $3.8\times10^{-5}$ & 0.03 \\
\hline \hline
\end{tabular} 
\caption{Nuclear spallation reactions for the production of $^{26}$Al. 
The reactions are grouped according to proton reactions (top), alpha
reactions (middle), and helion reactions (bottom). The second column
presents the cross section averaged over the normalized cosmic ray
spectrum. The relative abundances of the target nuclei
\citep{lodders2003} and the projectiles are listed in the last two
columns (see text). }
\label{table:react} 
\end{table} 

\bigskip 

This treatment uses the spallation cross sections provided by the
TENDL Nuclear Data Library \citep{koning2019}. The tabulated cross
sections listed in Table \ref{table:react} are averaged over the
(normalized) spectrum of cosmic rays, where we use $p=3$ as the
index. The resulting values are given in Table \ref{table:react}. Note
that the experimentally determined cross sections show significant
variations in their published values (compare these values with
\citealt{janisproton,janisalpha,janishelion}, and with the
weighted-averaged values given in \citealt{lee1998}\footnote{Note that  
this reference uses different weighting of the cross sections with
cosmic ray energy.}). The overall uncertainty in the cross sections is
less than a factor of two. Moreover, the values used here
\citep{koning2019} tend to be smaller than those of the other sources,
especially at higher cosmic ray energy, so that the estimates of this
paper represent lower limits. The abundances of the target nuclei are
assumed to have solar values (taken from \citealt{lodders2003}).
Finally, the relative fractions of projectiles $(p,\alpha,h)$ are
presented in the last column of the table (see Section
\ref{sec:produce}).

With the specifications given above, the total spallation cross
section for $^{26}$Al production defined by equation (\ref{sigtotal})
becomes $\sigmabar_T\approx740$ mb. It is useful to define the
benchmark value of abundance ratio 
\be
{\cal R}_{26}^{\textstyle \star} = 0.12
\left({\ndot\over3.75\times10^{34}\,\,{\rm s}^{-1}}\right)
\left( {\trunk \over 0.1 {\rm AU}} \right)^{-2} 
\left( {\sigmabar_T \over 740\,{\rm mb}} \right) \,. 
\label{benchmark} 
\ee
At the expected location $\trunk\sim0.1$ AU, this benchmark value of
the nuclear abundance ratio is much larger than the measured value in
solar system meteorites. However, this large ratio can only be
realized for moderate amounts of rocky material that is exposed to
particle radiation under optically thin conditions over long times.
In practice, energy losses within the rocky bodies reduces the
abundance ratio by a substantial factor. Departures from this
idealized limit are considered in the following section.

\begin{figure}
\includegraphics[scale=0.70]{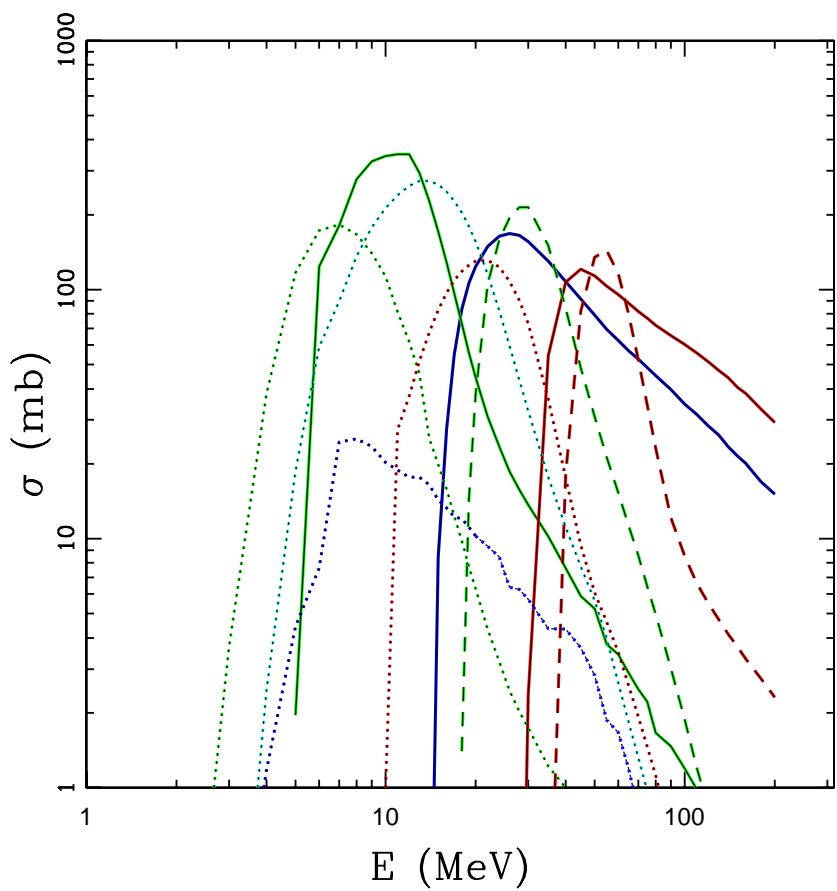}
\vskip-1.50truein
\caption{Cross sections for spallation reactions (from 
\citealt{koning2019}). The solid curves show the cross sections of
reactions with protons for target nuclei $^{26}$Mg (green), $^{27}$Al
(blue), and $^{28}$Si (red).  The dashed curves show the cross
sections of reactions with alpha particles for target nuclei $^{24}$Mg
(green) and $^{28}$Si (red).  The dotted curves show the cross
sections of reactions with heilons for target nuclei $^{24}$Mg
(green), $^{25}$Mg (cyan), $^{27}$Al (blue), and $^{28}$Si (red). }
\label{fig:cross} 
\end{figure}

\subsection{Optical Depth Considerations}  
\label{sec:opdepth} 

In general, the target nuclei could reside in a gaseous phase, be
locked up in dust grains, or be incoroporated in larger rocks.  Within
the current paradigm of planet formation, we expect most of the
targets of interest to condense onto dust grains or larger bodies. The
exposure of the targets to the cosmic ray flux is limited in two
stages. First, the cosmic ray flux can be attenuated on its path to
the rocky entities. Second, the cosmic ray flux is attenuated further
as it propagates within the rocky material. In the scenario considered
here, the plasma in the reconnection region is optically thin (see the
previous section) so that the first type of attenuation occurs through
interactions with the sea of rocky bodies. Once the cosmic ray flux
reaches the surface of a given rock, additional attentuation of the
cosmic rays takes place within the body itself.

Consider a total mass $M_R$ of rocky material distributed across 
the reconnection annulus, so that the column density of the region 
is given by 
\be
\Sigma_{\rm rock} = {M_R \over \pi \trunk^2} 
\approx \, 850 \,\,{\rm g} \,\,{\rm cm}^{-2} \,\,
\left( {M_R \over M_\oplus} \right) 
\left({\trunk \over 0.1 {\rm AU}}\right)^{-2} \,. 
\label{column} 
\ee
Here we consider the regime where the rocky bodies with radius $R$ are
partially optically thick. As a result, the constituent nuclei are
exposed to a fraction ${\cal F}$ of the cosmic ray flux that strikes
the surface, and a corresponding fraction $1-{\cal F}$ of the flux
is absorbed. Note that the fraction depends on the rock size, 
${\cal F} = {\cal F}(R)$, and this function is estimated below.  
The optical depth of the reconnection annulus is given by 
\be
\tau_{\rm rock} = \Sigma_{\rm rock} \,\,
{3 \over 4 \rho R} [1-{\cal F}(R)] \,,
\label{taurock} 
\ee
where $\rho\approx3.3$ g cm$^{-3}$ is the density of the rocks. As a
rough approximation, we can determine the fraction of the original
cosmic ray flux that is incident on the rocky bodies by using a 
simple one-dimensional slab geometry. This fraction is reduced 
further within the rocks, so that the nuclei are exposed to an 
overall fraction $f_\tau$ of the flux given by 
\be
f_\tau = {1 \over \tau_{\rm rock}} 
\left[ 1 - {\rm e}^{-\tau_{\rm rock}}\right] {\cal F}(R) \,.
\ee 
The first part of the expression accounts for the attenuation 
of the cosmic ray flux by the background sea of rocky material, 
before the flux reaches the surface of a given rock. The final 
factor ${\cal F}(R)$ accounts for the fraction of original surface 
flux that reaches nuclei within the rock. 

We thus need to estimate ${\cal F}(R)$. The cosmic rays entering the
rocky bodies lose energy in two ways: They can interact with nuclei
and drive nuclear reactions such as spallation. The interaction cross
sections for such processes are of order $\sigma\sim10^{-25}$
cm$^2$. They can also lose energy through Coulomb interactions with
electrons in the rock. In this case, the cosmic rays continually lose
energy and eventually come to a stop after encountering a sufficient
column density of material. For cosmic rays with incident energy
$E\sim10$ MeV, the required stopping column density $\Sigma_s\sim0.3$
g cm$^{-2}$ \citep{reedy1991,reedy2015}.  This stopping criterion
corresponds to an effective interaction cross section given by 
$\sigma_{\rm eff}\sim{A}m_p/\Sigma_s\sim10^{-22}$ cm$^2$, where $m_p$
is the proton mass, and $A$ is the mean atomic mass number of the
rocky material. Note that cosmic rays with higher energy have a larger
stopping column. Nonetheless, cosmic rays lose their energy within
rocks more readily than they interact with nuclei through spallation.

As cosmic rays propagate through the rocky material, the loss of 
energy is given by the expression
\be
{dE \over ds} = - F \left({E_{th} \over E}\right)^q \,,
\ee
where the power-law index $q\approx0.7$ and the coefficient $F$
depends on the composition of the target and the type of cosmic ray
particle \citep{reedy1991}. In addition, the coefficient is scaled
such that the benchmark energy $E_{th}$ is the threshold energy for
the reaction of interest. This loss equation can then be integrated to
find the energy of the particle remaining as a function of the
distance $s$ traveled, 
\be
E(s) = \left[ E_i^{q+1} - (q+1) F E_{th}^q s \right]^{1/(q+1)} \,.
\ee
Since the cosmic ray energy spectrum is much steeper than the energy
dependence $\sigma(E)$, we can make the approximation that the cross
sections are relatively constant above a threshold energy $E_{th}$.
In order for a cosmic ray to have sufficient energy to interact after
traveling a distance $s$, $E(s) > E_{th}$, and the initial energy
$E_i$ must be larger than the value
\be
E_i > E_{th} \left[ 1 + (q+1) (F/E_{th}) s \right]^{1/(q+1)} \,.
\ee
Suppose that a flux of particles $\Phi_0$ is incident on the rocky
surface. For the assumed cosmic ray energy spectrum from equation
(\ref{crspectrum}), the flux of particles with energy larger than 
a given energy $E$ can be written in the form 
\be
\Phi(>E) = \Phi_0 \left({E_0 \over E}\right)^{p-1} \,. 
\ee
The flux of particles above the threshold energy, evaluated at a 
distance $s$ into the rocky material, is thus given by 
\be
\Phi(s) = \Phi_0 \left({E_0 \over E_{th}}\right)^{p-1} 
\left[ 1 + (q+1) (F/E_{th}) s \right]^{-(p-1)/(q+1)} \,. 
\ee
The leading factor represents the incident flux of particles with
sufficient energy to induce spallation reactions. The factor in square
brackets determines how the flux is attenuated with the distance $s$
traveled through the rock.

Next we want to determine the effective cosmic ray flux impinging on
the target nuclei within a rocky body.  For simplicity, the rocks of
interest can be considered as uniform density spheres of radius $R$.
Within the reconnection region, the cosmic ray flux can be considered
as isotropic. This isotropic flux of particle is thus incident on a
spherical rock. For a given position within the rock $(r,\mu)$, where
$r$ is the spherical radial coordinate and the $\mu=\cos\theta$, the
distance coordinate $s$ is given by 
\be
s = r \mu + \left[ R^2 - r^2 (1-\mu^2) \right]^{1/2} \,. 
\ee
It is useful to make the following definitions, 
\be
\ell_0 = {E_{th} \over (q+1)F} \,, \qquad 
\gamma = {p-1 \over q+1} \,, \qquad {\rm and} \qquad 
\xi = {r \over R}\,,
\ee
where $\ell_0\approx0.1$ cm for $E_{th}=10$ MeV. The target nuclei
within the rock receive a only fraction ${\cal F}$ of the original
isotropic flux striking the surface, where 
\be
{\cal F} (R) = {3\over2} \left({\ell_0 \over R}\right)^\gamma 
\int_0^1 \xi^2 d\xi \int_{-1}^1 d\mu 
\left[(\ell_0/R) + \xi \mu + \left[ 1 - \xi^2 (1-\mu^2) \right]^{1/2} 
\right]^{-\gamma}\,\,.
\label{calf} 
\ee
In the limit of small rocky bodies $R\ll\ell_0$, the fraction 
${\cal F}\to1$ as expected. Moreover, the quantity $1-{\cal F}$
$\sim{\cal O}(R/\ell_0)$ in this regime. In the opposite limit
$R\gg\ell_0$, the fraction decreases rapidly with increasing $R$.  
In general, we must evaluate the integrals in equation (\ref{calf})
numerically.

\subsection{Nuclear Abundance Estimates} 
\label{sec:restimates} 

The considerations outlined above can be summarized by writing
the nuclear abundance ratio for $^{26}$Al in the form 
\be
{\cal R}_{26} = {\ndot \sigmabar_T \over \pi \trunk^2 \lambda} 
\,\left[ 1 - {\rm e}^{-\lambda\texp} \right]\,\,
\left[ {1- {\rm e}^{-\tau} \over \tau}\right]\, {\cal F}(R) \,. 
\label{alratio} 
\ee
The leading factor corresponds to the case where the target nuclei 
experience exposure from an unattenuated cosmic ray flux in the
long-term limit under optically thin conditions. The first
dimensionless correction factor in square brackets takes into account
departures from saturation due to shorter exposure times. The second
factor accounts for the attenuation of the cosmic ray flux before it
reaches the surfaces of the rocky bodies, where the optical depth of 
the reconnection region is given by $\tau\approx\tau_{\rm rock}$
(see equation [\ref{taurock}]). The final factor results from energy
loss of the cosmic flux within the rocky entities, primarily due to
Coulomb interactions, as approximated using the fraction ${\cal F}$
from equation (\ref{calf}).

\begin{figure}
\includegraphics[scale=0.70]{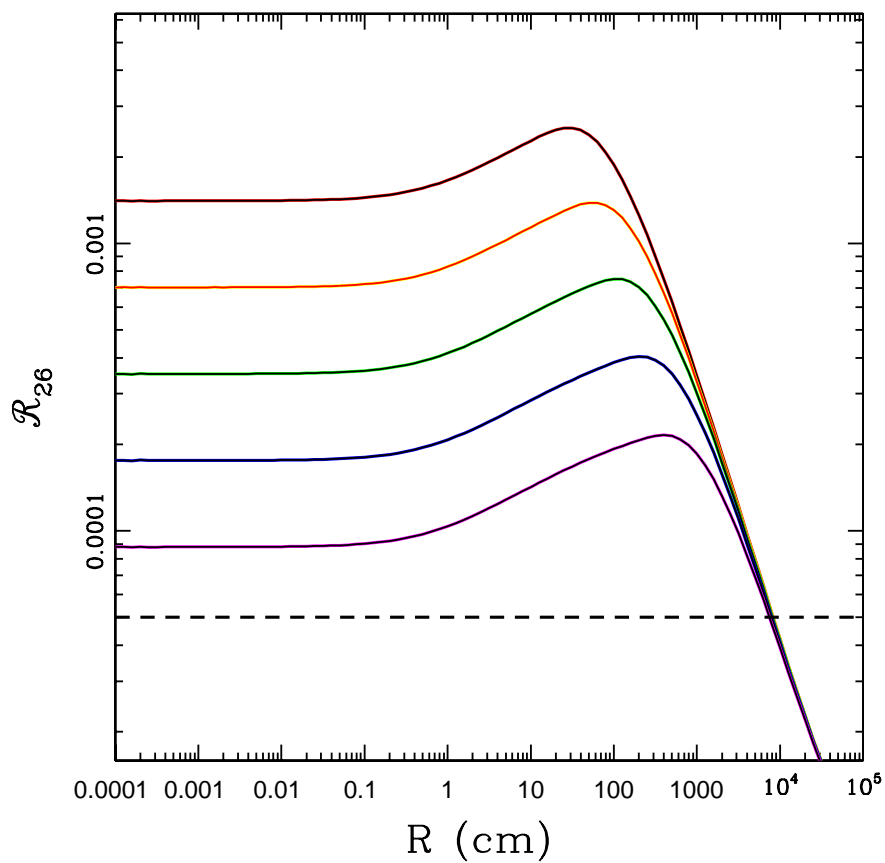}
\vskip-1.50truein
\caption{Estimated abundance ratio ${\cal R}_{26}$ (for  
$^{26}$Al/$^{27}$Al) as a function of radius of rocky bodies.  
Curves are shown for different amounts of rocky material in the
reconnection annulus, corresponding to increasing mass from top to
bottom: $M_R/M_\oplus$ = 1/2 (red), 1 (orange), 2 (green), 4 (blue), 
and 8 (magenta). The lower dashed line marks the abundance ratio 
inferred for our solar system. } 
\label{fig:ratio} 
\end{figure}

The resulting estimates for the abundance ratio ${\cal R}_{26}$ are
shown in Figure \ref{fig:ratio} as a function of the radius $R$ of the
rocky bodies. Results are shown for a range of surface densities
$\Sigma_{\rm rock}$, which sets the level of the optical depth and
depends on the total mass of rocky material in the reconnection region
(see equation [\ref{column}]).  Curves are shown for total masses
corresponding to the planet types of interest, specifically
$M_R=0.5M_\oplus$ (upper red curve), $1M_\oplus$ (orange), $2M_\oplus$
(green), $4M_\oplus$ (blue), and $8M_\oplus$ (lower magenta curve).
These estimates are calculated in the long-time limit so that
$\lambda\texp\gg1$, which is valid for exposure times of order 1 Myr
or longer. For comparison, the lower dashed line shows the inferred
abundance ratio for our solar system. Although the estimated abundance
ratios are substantially larger than those of our solar system, they
are also much smaller than the maximum benchmark value given by
equation (\ref{benchmark}).

In the limit where the rocky bodies are small, the curves in Figure
\ref{fig:ratio} are flat and depend only on the total optical depth of
the reconnection region (see equation [\ref{taurock}]). In this
regime, the quantity $(1-{\cal F})/R\to$ {\sl constant}
$\sim1/\ell_0$, so attenuation of the cosmic rays depends only on the
total amount of material (independent of $R$).  Notice also that most
of the attenuation results from Coulomb effects, rather than nuclear
reactions. As the radius $R$ increases, the optical depth 
$\tau_{\rm rock}$ of the background rocky material decreases. The
reconnection region becomes optically thin for rock sizes $R\gta200$
cm $(M_R/M_\oplus)$.  At the same time, the optical depth of each
individual rock increases, with an increasing fraction of its interior
shielded from the incident cosmic ray flux. The competition between
these two effects results in the broad maximum shown for $R=50-100$
cm. For even larger rock radii, the abundance decreases sharply as an
increasing fraction of the constituent nuclei are shielded. The
enrichment levels are less than solar system values for rock sizes
$R\gta10^4$ cm, and become negligible for $R\gta1$ km. 

For small planets, $\mplan\sim1M_\oplus$, Figure \ref{fig:ratio}  
indicates that enrichment levels can be more than an order of
magnitude larger than those inferred for the Solar System
(equivalently, a maximum mass of $\sim15M_\oplus$ of material can be
enriched to Solar System values).  In previous studies, which focus on
meteoritic enrichment, the exposure times are often smaller than those
considered here, with correspondingly smaller yields (see, e.g.,
\citealt{shu1996} to \citealt{sossi2017}). Such studies also require
the enriched material to be carried out to larger heliocentric
distances.  This scenario --- which allows for rocky material to be
exposed to the largest cosmic ray fluxes over the longest times --
thus maximizes local enrichment. Nonetheless, given the finite number
of energetic cosmic rays, there is a limit on the amount of material
that can achieve such high abundances of $^{26}$Al. Previous studies
estimate that only $\sim3M_\oplus$ of material can reach solar system
abundances \citep{duprat2007}, where this limit can be extended up to
$\sim10M_\oplus$ under favorable circumstance \citep{fitoussi2008}.
These results are comparable to the mass of $\sim15M_\oplus$ found
here (where we assume long exposure times, so that enrichment is
saturated, and use a somewhat different treatment for Coulomb
losses). Significantly, all of these estimates ($3-15M_\oplus$) fall
below the estimated rocky inventory of the Solar System ($40-80$
$M_\oplus$). This finding suggests that spallation cannot enrich the
entire early solar nebula to the high isotopic abundances observed in
meteorites. 

For typical pebble sizes $R\sim$ few cm, the estimated abundance
ratio has the approximate value ${\cal R}_{26}\approx$ 0.001
$(M_R/M_\oplus)^{-1}$. For a planet with mass $\mplan\approx M_R$ = 1
$M_\oplus$, for example, this enrichment level is $\sim15$ times larger 
than that inferred for our solar system. Note that smaller bodies 
($R\ll1$ cm) are likely to be swept up into the accretion flow,
deposited onto the stellar surface, and be unavailable for planet
formation.\footnote{For example, \cite{gounelle2001} use the size 
range $R\sim100 \mu$m -- 10 cm for the cores of enriched rocks in 
our solar system.}  The key issue is thus the length of time $\texp$
that the pebbles remain small ($\sim1$ cm) before being incorporated
into larger planetesimals. Next we note that radioactive heating
rates, and other physical quantities of interest, depend on the mass
fraction $X_{26}$ of $^{26}$Al, where 
\be 
X_{26} = X_{27} {\cal R}_{26} \approx 4.5 \times 10^{-6} 
\left( {M_R \over M_\oplus}\right)^{-1} \,.  
\ee
Here we have taken $X_{26}$ and $X_{27}$ to be the mass fractions of
the rocky material, and have used $X_{27}\approx4.5\times10^{-3}$ for
solar system material (estimated from Table 2 of \citealt{lodders2003}).  
Another quantity of interest is the power per unit volume $\heat$ due
to radioactivity, given by 
\be 
\heat = \,\lambda\,\epsilon\,X_{26}\,
\left({\rho \over 26m_p}\right)\,{\rm e}^{-\lambda t} \, 
\approx 0.066 \,{\rm erg}\,{\rm s}^{-1}\,{\rm cm}^{-3}\,
\left({M_R\over1M_\oplus}\right)^{-1}\,{\rm e}^{-\lambda t} \,,
\ee 
where $\epsilon\approx4$ MeV is the energy deposited per decay of the
$^{26}$Al nucleus. The second approximate equality evaluates the
volume heating rate for standard values of the parameters.

Given these estimates for radioactive enrichment levels, it is useful
to assess how the results depend on the input parameters and other
uncertainties.  Figure \ref{fig:ratio} shows how the predicted
$^{26}$Al abundances vary with the size of the rocky bodies and the
total optical depth of the reconnection region (given by the total
mass in solids).  These curves are calculated in the regime of long
exposure times ($\lambda t_{\rm exp}\gg1$); shorter exposure times can
be considered by including the dimensionless correction factor in
equation (\ref{alratio}). Additional input to the estimates include
the reaction cross sections, the target abundances, the relative
populations of cosmic ray species, and the power-law index of the
particle radiation spectrum. All of these variables combine to specify
the net cross section $\sigmabar_T$ (see equation [\ref{ratone}]).
The cross sections for individual reactions $\sigmabar_k$ averaged
over the cosmic ray spectrum show little variation with the spectral
index because the cross sections are relatively localized in energy
(\citealt{duprat2007,fitoussi2008}; see also Figure \ref{fig:cross}). 
The target abundances scale with the metallicity (where solar values
are used here). One uncertain quantity is the fraction of the cosmic
ray flux in helions, taken here to be 0.03. This fraction could vary
by factors of $\sim3$, but the helion contribution to the overall
production rate is only $10.7\%$, so that the uncertainty in the 
$^{26}$Al yield is also $\sim10\%$. Perhaps the largest uncertainty is
the net luminosity in energetic particles. The value used here,
$\crlum=10^{-4}L_\ast$, provides a good estimate for the expectation
value \citep{feigelson}; however, the system to system scatter allows
for an order of magnitude variation from the mean, with a
corresponding variation in the expected radioactive yields. 

\section{Short-Lived Radionuclides in Planetesimals and Planets} 
\label{sec:effects} 

Given the elevated levels of $^{26}$Al enrichment expected for rocky
bodies residing in the reconnection region, this section explores
their effects on newly formed planetesimals and fully formed planets.
Here we assume that small ($\sim1$ cm) bodies (pebbles) are irradiated
by cosmic rays and attain a sizable inventory of radioactive nuclei.
Streaming instability, or some alternate process, then produces much
larger bodies (planetesimals) on a relatively short time scale. These
larger entities, with sizes 10 -- 100 km, are heated by radioactive
decay. Here we estimate the relevant time scales and determine the
circumstances under which these bodies can melt. If the planetesimals
are readily incorporated into full-sized planets, then the planets 
themselves can have a subsantial internal luminosity. 

\subsection{Equilibrium Time Scale} 
\label{sec:timescale} 

The time scale over which a planetesimal adjusts its internal thermal
structure is governed by the Kelvin-Helmholtz time, i.e., the time
required for the internal luminosity to radiate away the heat content
of the planetesimal. The luminosity is given by $L_{\rm in}$ = 
$\heat M/\rho$, where $M$ is the mass of the rocky body. The heat
content is given by ${\cal U}$ = $CMT$, where $C$ is the specific heat
capacity of the rock.  The characteristic cooling time scale is thus
given by 
\be
t_{\rm cool} = { {\cal U} \over L_{\rm in}} = {C\rho T\over\heat} 
\approx 27,000\,{\rm yr} 
\left({\heat\over 0.066 \,{\rm erg}\,{\rm s}^{-1}\,{\rm cm}^{-3}\,}\right)^{-1} 
\left({T\over2000\,{\rm K}}\right)\,. 
\ee
The temperature reference scale is chosen because rock melts when
$T\gta2000$ K.  The resulting cooling time is substantially shorter
than the decay time scale for the SLRs ($\sim1$ Myr). Note that this
time scale is independent of the mass of the planetesimal.

\subsection{Size Scale for Planetesimal Melting}  
\label{sec:psmelting} 

The internal radioactive luminosity of the rocky body $L_{\rm in}$
must be transported outwards, so that the temperature obeys a
diffusion equation of the general form 
\be
\rho C {\partial T \over \partial t} = {1 \over r^2} 
{\partial \over \partial r} 
\left( \conduct r^2 {\partial T \over \partial r} \right) + 
\heat(r,t) \,, 
\label{tempdiff} 
\ee
where $\heat(r,t)$ is the heating due radioactivity and $\conduct$ is
the thermal conductivity (see \citealt{hevey2006} and references
therein). We can divide out the density $\rho$ and the specific heat
$C$ to obtain 
\be
{\partial T \over \partial t} = {\kappa \over r^2} 
{\partial \over \partial r} 
\left( r^2 {\partial T \over \partial r} \right) + 
{\heat_0 \kappa \over \conduct} {\rm e}^{-\lambda t} \,, 
\label{tempdiff2} 
\ee
where $\heat_0$ is a constant and $\kappa\equiv{\conduct}/C\rho$ is
the thermal diffusivity of the rocky material.  If we assume that the
initial temperature of the planetesimal is that of the ambient region
at $T_0$, then the initial condition $T(r,t=0)$ = $T_0$. Following
standard treatments of this problem \citep{carlslaw}, the solution has
the form
\be
T(r,t) = T_0 + {\heat_0 \kappa \over \lambda \conduct} {\rm e}^{-\lambda t} 
\left[ {R \sin qr \over r \sin qR} - 1 \right] + 
{\heat_0 R^2 \over \conduct} {R \over r} \sum_{n=1}^\infty 
a_n \sin\left({n\pi r\over R}\right)  
{\rm e}^{-\kappa n^2 \pi^2 t/R^2}\,,
\label{tsolution} 
\ee
where we have defined 
\be
q \equiv \sqrt{\lambda \over \kappa} \qquad {\rm and} \qquad 
a_n \equiv {2 \over \pi^3} 
{(-1)^n \over n (n^2 - \lambda R^2/\kappa\pi^2)} \,. 
\ee
The analytic solution (\ref{tsolution}) assumes that the thermal 
conductivity and diffusivity of the rock is constant. However, once 
the temperature is high enough to melt the rock, $T\gta2000$ K, the 
molten body can transport heat via convection and the thermal 
properties change. 

\begin{figure}
\includegraphics[scale=0.70]{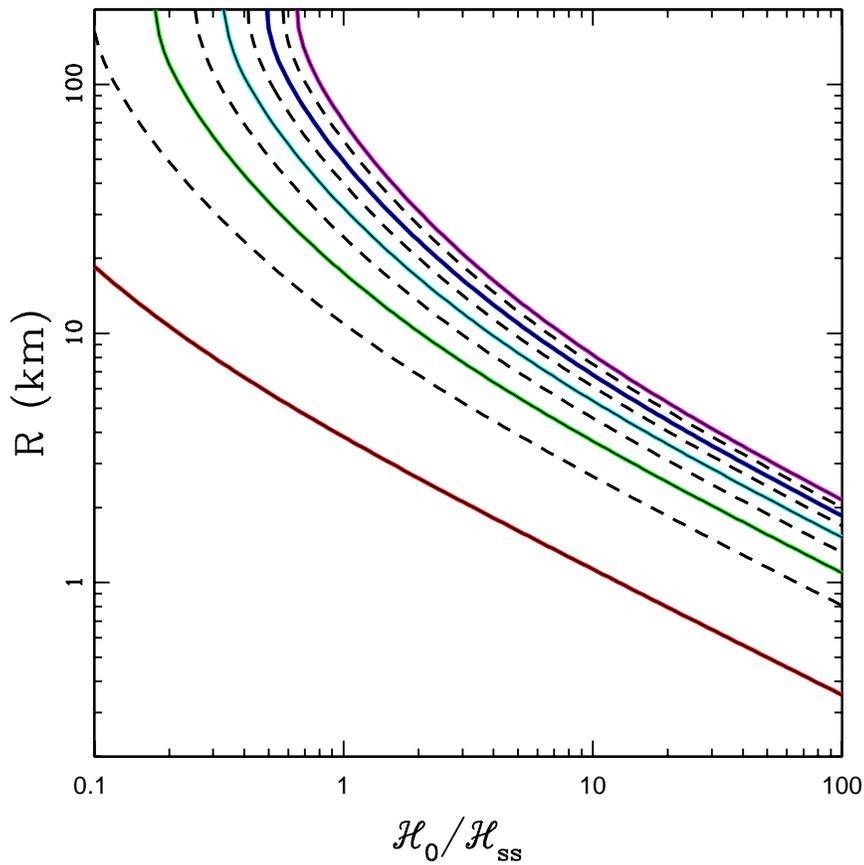}
\vskip-1.50truein
\caption{Internal temperatures of planetesimals as a function of 
size $R$ and radioactive heating rate $\heat_0$. The curves show 
contours for which the planetesimals reach a given temperature over 
90 percent of their radial extent, for $T$ =1000 K (red), 2000 K (green), 
3000 K (cyan), 4000 K (blue), and 5000 K (magenta). Dashed curves show 
$T$ = {\sl constant} contours for intermediate values of temperature. } 
\label{fig:hrplane}
\end{figure}

Here we use the solution (\ref{tsolution}) to map out the parameter
space for which planetesimals can melt. For a given value of the
radioactive heating rate $\heat_0$, we find the minimum planetesimal
size $R$ that is required to achieve a given temperature.  For the
sake of definiteness, we measure the heating rate in terms of the 
benchmark value $\heat_{SS}=6.4\times10^{-3}$ erg s$^{-1}$
cm$^{-3}$ that is often used for planetesimals in our solar
system.\footnote{Note that the value $\heat_{SS}$ uses the inferred
  ratio ${\cal R}_{26}$ for the solar system, but assumes an aluminum
  rich composition (see \citealt{hevey2006}).}  In addition, we
require that 90 percent of the radial extent of the planetesimal reach
the given temperature.  Figure \ref{fig:hrplane} shows the resulting
contours of constant temperature in the plane of size $R$ versus
heating rate $\heat_0$.  The curves correspond to temperatures from $T$
= 1000 K (red) up to $T$ = 5000 K (magenta), with the dashed curves
showing intermediate temperature values. The expected temperature
required for melting of rock is about 2000 K, as depicted by the green
curve in the Figure.  As a result, planetesimals are expected to
become almost fully molten for the parameter space above the green
curve.

For the relatively large radioactive heating rates of interest,
$t_{\rm cool} < t_{1/2}$, so that the temperature structure of the
planetesimal can reach an equilibrium state.  As a result, the
steady-state solution of the diffusion equation (\ref{tempdiff}) is
applicable and takes the form 
\be
T(r) = T_S + T_x \left(1 - {r^2 \over R^2}\right) \,,
\ee
where $T_S$ is the surface temperature and where the temperature 
scale $T_x$ is determined by the rock properties
\be
T_x \equiv {\heat R^2 \over 6\conduct} \approx 51 \, {\rm K} 
\left({\heat \over \heat_{SS}}\right) 
\left({R \over 1\,{\rm km}}\right)^2\,. 
\ee
In order for most of the planetesimal volume to be molten, the
temperature scale $T_x$ must be significantly above the melting
temperature of rock. For example, if we require $T>2000$ K at
$r/R=0.90$, the scale $T_x\sim5000$ K. For solar system heating rates,
one thus requires $R\gta10$ km.  As the SLR abundances increase, the
critical radius for melting scales (approximately) as
$R_{crit}\sim\heat^{-1/2}$ (consistent with the slope of the curves 
on the right side of Figure \ref{fig:hrplane}). 

\subsection{Internal Luminosity of Rocky Planets} 
\label{sec:luminosity} 

Once a planet has formed out of the planetesimals, the internal
luminosity will be enhanced due to radioactive decay of $^{26}$Al (and
other SLRs). This luminosity contribution can be written in the form 
\be
L_{\rm in} = \lambda \epsilon X_{26} 
\left({\mplan\over 26m_p}\right) \, {\rm e}^{-\lambda t}
\approx 1.2 \times 10^{26}\,{\rm erg}\,{\rm s}^{-1}\, 
\left({\mplan\over M_R}\right) {\rm e}^{-\lambda t} \,.
\ee
Here the time variable is measured from the production of the
planetesimals, as radioactive enrichment becomes highly inefficient
when rocky bodies reach that size.  Since the planet, with mass
$\mplan$, is made from the rocky material in the reconnection annulus,
with mass $M_R$, we expect the two mass scales to be comparable. As a
result, this scenario defines a characteristic planetary luminosity
$L\sim3\times10^{-8}L_\odot$, which is independent of the planetary
mass (if planet formation is efficient so that $\mplan\sim M_R$). 

For comparison, the power intercepted by the planet from its
host star takes the form 
\be
P_\ast = L_\ast \left({\rplan \over 2r}\right)^2 \approx 
1.8 \times 10^{26} \,{\rm erg}\,{\rm s}^{-1}\, 
\left({\rplan \over R_\oplus}\right)^2
\left({r \over 0.1{\rm AU}}\right)^{-2} \,, 
\ee
where we have used $L_\ast=L_\odot$.  For earth-sized planets, the
external power is roughly comparable to the internal power for the
expected abundances of SLRs. For larger planets, the external power is
relatively larger. This ordering only holds for newly formed planets,
however, as the internal luminosity decreases exponentially with time.
Finally, note that the residual heat left over from planet formation 
provides another power source for young planets. 

\section{Conclusion} 
\label{sec:conclude} 

This paper constructs a model for the production of $\,^{26}$Al, and
other SLRs, during the epoch of planet formation.  In this scenario,
cosmic rays are generated via magnetic reconnection events near the
truncation radius that marks the inner edge of the planet-forming
disk. The rocky bodies in this region have typical sizes $R\sim1$ cm,
and are exposed to an intense flux of cosmic rays that drive
spallation reactions. The enriched pebbles are subsequently
incorporated into planetesimals ($R\sim1-100$ km), which eventually
become fully formed rocky planets ($R\sim6000-10^4$ km). Our main
results are summarized below (Section \ref{sec:results}) along with a
brief discussion of their implications (Section \ref{sec:discuss}).

\subsection{Summary of Results}  
\label{sec:results} 

The magnetic truncation radius of T Tauri star/disk systems falls in a
range centered on $\trunk\sim0.1$ AU. This region is coincident with a
substantial population of observed exoplanets found in multi-planet
systems (Figure \ref{fig:rmag}).

The planet-forming material in the reconnection region is exposed to
intense cosmic ray radiation that will synthesize radioactive nuclei
through spallation reactions. The resulting nuclear isotope ratio for
$^{26}$Al has a fiducial value ${\cal R}_{26}\approx10^{-3}$, a factor
of $\sim20$ larger than the isotope ratio inferred for our solar
system (Figure \ref{fig:ratio}).

The predicted isotope ratio ${\cal R}_{26}$ scales inversely with the
total amount of rocky material $M_R$ in the reconnection region. If we
assume that the planets form from the same reservoir of material, then
${\cal R}_{26}\propto\mplan^{-1}$.  The cosmic radiation is attenuated
by propagating through rocky material, where Coulomb losses provide
the dominant source of energy loss. This treatment does not include
spallation produced via secondary particles (e.g., \citealt{pavlov}), 
so the abundances presented here represent lower limits.

With the enhanced abundances of SLRs, the minimum radius necessary for
planetesimals to become completely molten decreases (Figure
\ref{fig:hrplane}). For the expected isotope ratio corresponding to
$M_R=1M_\oplus$, planetesimals are predicted to melt for radii greater
than $R\gta3$ km. For comparison, the radius needed for planetesimals
to melt is about $\sim20$ km for the standard abundance of radioactive
material.

Since the planetesimals are expected to entirely molten, the
composition of planets forming near $r\sim0.1$ AU can be different
than those forming farther out. The volatile component of the planet
will be determined by those materials that are soluable in magma (see
the review of \citealt{chao2020}). In particular, water is efficiently
out-gassed in molten planetesimals (e.g., \citealt{lichtenberg}) so
that planets forming in the reconnection region are predicted to be
severely dehydrated.  These effects should be eludicated in future
work.

With the expected enhancement of SLRs, planets with masses
$\mplan\sim1M_\oplus$ generate significant internal luminosities due
to radioactive decay. If the planets form rapidly, this internal
luminosity can be larger than the power intercepted from the host
stars (at $r=0.1$ AU and at early times). As a result, assessments of
atmospheric evaporation must include internal energy sources due to
radioactivity.

For the most common stars in the galaxy, with masses $M_\ast\sim0.25$
$M_\odot$, the reconnection region for cosmic ray exposure
($\trunk\sim0.1$ AU) corresponds to the habitable zone (after the
stars reach the main sequence).  This paper shows that the raw
materials that make up this class of potentially habitable planets are
naturally irradiated by local cosmic rays (Figure \ref{fig:rmag}),
leading to spallation and radioactive enrichment (Figure
\ref{fig:ratio}).  One potential consequence of this chain of events
is that habitable zone planets orbiting red dwarfs are depleted in
water and other volatiles.

\subsection{Discussion} 
\label{sec:discuss} 

For the scenario developed in this paper, the predictions for enhanced
levels of radioactive enrichment depend on the timing of
planet-forming events. The rocky bodies --- in the form of pebbles ---
must be large enough ($R\gta1$ cm) that they can reside in the
reconnection region and avoid being swept into the star (keep in mind
that this region is depleted in gas). At the same time, the rocks must
remain small small enough ($R\lta300$ cm) so that they can be
efficiently irradiated and enriched.  The subsequent production of
planetesimals must then occur rapidly in order for the radioactive
species to remain active and melt the larger rocky bodies. The
exposure time $\texp$ of the pebbles must thus be somewhat longer than
the half-life $t_{1/2}\sim0.72$ Myr, whereas the time $t_{\rm pls}$
required to produce planetesimals must be shorter. The required
ordering of time scales ($t_{\rm pls}<t_{1/2}<\texp$) is typically
satisfied in planet formation models that operate via the streaming
instability \citep{drazkowska,simon,johansen}. However, if departures
occur, for example shorter exposure times while the rocky bodies are
small, the level of radioactive enrichment would be correspondingly
smaller. A related issue is that the gas in the reconnection 
must be optically thin ($\Sigma\lta0.01$ g cm$^{-2}$) for maximum
exposure to cosmic rays, whereas the streaming instability is most
efficient for comparable densities of gas and solids
\citep{youdin2005}.  Since the surface density of solids is larger in
this scenario, the growth of planetesimals by streaming instability is
correspondingly slower.  On the other hand, the solids are already
concentrated and can collapse to form planetesimals via gravitational
instability \citep{goldward,youdin2002}. 

Another possible way to reduce enrichment is for planets to form at
more distant radial locations in the disk and then migrate inwards.
Since the degree of radioactive heating can be much larger for the
case of in situ formation, the composition of planets, and the
corresponding the mass-radius relation, could be different for the two
scenarios. Future work should thus outline the detailed chemical
compositions resulting from melted planetesimals with varying levels
of radioactive enrichment. Such studies could then predict how
planetary properties are different for planets that migrate versus
those that form locally.

In the present treatment, the solution to the diffusion equation for
the planetesimal temperature assumes that the body is made of solid
rock (so that the thermal diffusivity is constant).  The resulting
temperature profiles provide a good estimate for when and where the
rocky planetesimals become hot enough to melt.  After the body is
sufficiently molten, however, convection carries the heat through the
planetary layers where temperatures are high. Future work should thus
construct evolutionary models (building on the work of
\citealt{hevey2006}) for planetesimals where the internal heat is
transported convectively. The time dependent temperature structure can
then be used to determine the degree of volatile losses and hence the
chemical composition of the objects. In addition, the largely molten
nature of the planetesimals could affect their subsequent assembly
into larger planetary bodies. Collisions between the resulting liquid
entities should thus be studied.  Once planets are formed, the
internal luminosity from SLRs will affect the surface temperatures and
the rate of atmospheric mass loss. This process should also be modeled
in greater detail.

This paper has focused on the isotope $^{26}$Al, which is expected to
have the largest impact on planetesimals and forming planets.  Other
SLRs can be made through spallation and should be included in future
work. Consider, for example, the case of $^{53}$Mg, with half-life 3.7
Myr. The solar system abundance of this species is a factor of 10
smaller than that of $^{26}$Al. If the two isotopes are produced with
initial abundance ratios similar to those of our solar system, then 
the relative power contribution will vary with time according to
$L_{\rm Mn53}/L_{\rm Al26}$ = $(\lambda_{53}/\lambda_{26})$
$(X_{26}/X_{53})$ $\exp[-(\lambda_{26}-\lambda_{53})t]$ $\approx0.02$
$\exp[t/1.3{\rm Myr}]$. The manganese luminosity should thus dominate
after ages of $\sim5$ Myr, although by that time the overall power
will have decreased by a factor of $\sim100$. In addition to SLRs,
long-lived radioactive nuclei can also be produced by spallation. For
example, protons of sufficiently high energy ($\sim100$ MeV) can
produce $^{40}$K through spallation reactions with iron (mostly
$^{56}$Fe, which is relatively abundant).

This paper shows that planets forming in close proximity to their host
stars are exposed to enormous fluxes of particle radiation and can
become substantially enriched in SLRs. The degree of radioactive
enrichment will vary with the planet location, the timing of planet
forming events, and the details of the local magnetic field structure.
The resulting variations in planet properties --- due to heating,
differentiation, out-gassing, evaporation, and other processes ---
will thus add to the diversity found in the observed population of
extra-solar planets.


\bigskip 

\noindent
{\bf Acknowledgment:} We would like to thank Konstantin Batygin,
Juliette Becker, George Fuller, Evan Grohs, Alex Howe, and Chris
Spalding for useful discussions. We also thank an anonymous referee
for comments and suggestions.  This work was supported by the
Leinweber Center for Theoretical Physics and by the University of
Michigan.

\end{document}